%% file: Main-0803-2026.tex
\documentclass{article}

\input{MyMacro}

\title{A new estimation for probability distributions using $L_{2}$ optimization}
\author{Jiwoong Kim\\
University of South Florida}

\usepackage{setspace}

\begin{document}
\maketitle
\begin{abstract}
Many probability distributions belong to a location-scale family. One such an example is a normal distribution. Probability distributions of the family possess many desirable merits and have colossal amount of real-world application. In this paper, we propose a novel method to estimate both location and scale parameters of probability distributions and investigate its asymptotic properties. After examining theoretical aspects of the proposed estimators, we will empirically demonstrate that they compares favorably with other estimators through simulation studies.
\end{abstract}
\noi

\section{Introduction}
In statistics and probability, a location-scale family is a collection of probability distributions that contains two parameters that govern the overall shape of distributions such as symmetry and dispersion. The most well-known probability distribution belong to the family is a normal distribution, and its location and scale parameters -- commonly denoted by $\mu$ and $\si$ in the statistical literature -- convey information about the distribution; the distribution is centered at the location parameter, while its dispersion can be expressed as the scale parameter. A special case of a location-scale family is a scale family; as the name implies, scale family distributions contain a scale parameter only, and
and the most famous probability distribution of the family is an exponential distribution.

The quintessence of probability distributions belonging to a location-scale family is that they can be \textit{standardized} by shifting their centers and scaling up or down their dispersion, which simplifies statistical modeling when they are fitted to the real datasets. This property is, sometimes, regarded as the raison d'etre of a location-scale family. For example, to make the statistical inference about some normally distributed data, we solely need a standard normal distribution whose location and scale parameters are 0 and 1, respectively.

Reflecting the importance of the location-scale families, many methods have been suggested to estimate both location and scale parameters; the most popular one is definitely the maximum likelihood (ML) estimation. However, the ML estimator is known to perform very poorly in the presence of outliers. Since real data often contains outliers, it is worth seeking an alternative estimator. It is against this backdrop of the need for a robust estimator that we started this study. To this end, we will propose a minimum distance (MD) estimator and demonstrate that it possesses the desirable robustness; the asymptotic property is a by-product of the investigation. The proposed MD method differs from traditional methods in the literature in that it parameterizes the parameter of interest in a more computationally friendly way, thereby reducing the computational burden of estimation; this is the main contribution of this article to the literature on MD estimation.

The rest of this article is organized as follows. Section \ref{Sec:MDE} introduces the MD estimation and discusses the resulting estimator, including the literature review of the proposed method and its asymptotic properties. Section \ref{sec:desirable_propoerties} discusses the other desirable properties of the proposed estimator. Section \ref{Sec:simulation_studies} compares the proposed method with other methods through simulated experiments and reports the empirical findings. Section \ref{Sec:conclusion} concludes the article by suggesting potential future research that extends from this study.

\section{Minimum distance estimation}\label{Sec:MDE}
\subsection{Literature review}
During the 1970s and 1980s, many statisticians -- e.g., \cite{Koul1970}, \cite{Millar1984}, \cite{Donoho88a}, and \cite{Donoho88b} -- have conducted research on the MD estimation since it was proposed by \cite{Wolfowitz1953}: see also references in \cite{Koul2002}. The distance function measures the discrepancy between the observed random sample and the assumptions underlying the theories. More specifically, the distance function –- which contains the parameter of interest as an argument -- computes the numeric difference between the empirical function and the modeled function, constructed from observed data and assumptions, respectively. Then, the MD estimation method, as the name implies, seeks the optimal value that minimizes the distance function. Researchers investigated the resulting estimator after employing various distance functions. For example, \cite{Beran} used the Hellinger distance using empirical and modeled density functions. Among many research works on the distance function in the literature of the MD estimation, \cite{Parr} demonstrated that the MD estimator obtained from the Cramer von-Mises (CvM) distance function exhibits better robustness than those obtained from other distance functions.

In the past two decades, however, no more rigorous research has been conducted; only a few studies have further investigated MD estimation. \cite{Kim2018} proposed a novel algorithm to compute the MD estimator, while \cite{Kim2020} demonstrated that the MD estimator maintains the desirable asymptotic properties under the assumption of independent observations even when independence doesn’t hold. Applying the MD method to a discrete distribution, \cite{Kim2026} demonstrated that the MD estimator still retains asymptotic properties and robustness, thereby comparing favorably with other well-celebrated estimators, including the ML estimator.

One of the fundamental reasons the popularity of the MD method has been waning rapidly is the complexity of its distance function. The empirical distribution function, which is a collection of indicator functions of the observed sample and the parameter of interest, is the main culprit obstructing the search for the optimal solution to the distance function. Since the indicator function is not smooth, it is not differentiable with respect to the parameter; unlike the ML estimation, the closed-form expression for the solution does not exist. Therefore, research on MD estimation should rely on computationally expensive numerical methods to solve the optimization problem. \cite{Kim2026} modified the distance function so that the empirical distribution function contains only the observed sample, while the modeled distribution function contains the parameter of interest. As a result, the distance function became smooth and differentiable with respect to the parameter, thereby reducing the computational cost; see, e.g., Figure \ref{fig:distant_function}. Findings in \cite{Kim2026}, however, are limited in that the modified distance function is valid only for a discrete probability distribution, the estimation problem of which is presumed to be less challenging than that of a continuous probability distribution function. In this article, we extend his approach to estimating the rate parameter of the exponential distribution.

\subsection{CvM-type distance function}\label{Sec:CvM_distance_function}
Consider a sample of random observations, $\vX$ that are independently, identically distributed (iid). Let $f(\cdot;\bth_{0})$ and $F(\cdot;\bth_{0})$ denote the density and distribution functions with a vector of location and scale parameters $\bth_{0}=(\mu_{0}, \si_{0})'\in \mathbb{R}^{2}$. Note that the partial derivatives of the distribution function $F$ with respect to its location and scale parameters are
\ben\label{eq:partial_derivative_F}
\frac{\partial F(y;\bth)}{\partial \mu}=-f(y;\bth)\,\,\textrm{ and }\,\, \frac{\partial F(y;\bth)}{\partial \sigma}=-\frac{(y-\mu)}{\si}f(y;\bth),
\een
respectively. To estimate the parameter $\bth_{0}$, we first define the distance function
\ben\label{eq:dist_fct}
\cL(\bth) = \int\left[ \sti d_{ni}\Big\{\textrm{I}(X_{i}\le y)-F(y;\bth) \Big\} \right]^{2}\,dH(y),
\een
where $\textrm{I}(\cdot)$ is an indicator function,  $d_{ni}\in \mathbb{R},\,1\le i\le n$ are real numbers, and $H(y)$ is an integrating measure; $d_{ni}$'s can uniformly be $1/\sqrt{n}$ as done in Kim (2026), and $H$ can simply be the Lebesgue measure. Note that traditional MD methods place $\bth$ inside the indicator function and use a normalized distribution function, which causes the distance function to be neither smooth nor differentiable with respect to $\bth$, thereby rendering the optimization problem defined in (\ref{eq:opt}) more challenging and incurring a high computational cost. Thus, parameterizing the distribution function rather than the indicator function will facilitate faster computation of the MD estimator.

Next, we define the MD estimator
\ben\label{eq:opt}
\cL(\widehat{\bth}) = \inf_{\bth\in \mR^{p}}\cL(\bth).
\een
When the ML estimation was introduced to the literature of statistics and probability, the asymptotic normality of the resulting ML estimator adds cachet to its popularity. Based on the asymptotic normality of the ML estimator, popular statistical inferences, such as hypothesis tests and confidence intervals, become possible as the sample size increases. Fortunately, such an asymptotic property is also viable for the MD estimation, provided that it satisfies \textit{uniformly locally asymptotically quadratic} (ULAQ) conditions. To begin with, we state reproduce the ULAQ conditions from \cite{Koul2002}. Define the neighborhood $\cN_{b}(\bth_{0}):=\{\bth\in\mathbb{R}^{p}:\|\D_{n}(\bth-\bth_{0})\|\le b\}$ where $\D_{n}$ is a $2\times 2$ matrix and satisfies the assumption \tbf{(a.2)} below.
\begin{itemize}
  \item[\textbf{(u.1)}] There exists a random variable $\bSn(\bth_{0})$ and a real number $\bW_{n}(\bth_{0})$ such that for all $0<b<\iny$
   \benn
   \sup_{\bth\in \cN_{b}(\bth_{0})}\left|\cL(\bth)- \cL(\bth_{0})-2(\bth-\bth_{0})'\bSn(\bth_{0})-(\bth-\bth_{0})'\bW_{n}(\bth_{0})(\bth-\bth_{0})\right|=o_{p}(1).
   \eenn
  \item[\textbf{(u.2)}] For all $\vep>0$, there is a $0<z_{\vep}<\iny$ such that
  \benn
    \mathbb{P}\big(|\cL(\bth_{0})|\leq z_{\vep} \big)\geq 1-\vep.
  \eenn
  \item[\textbf{(u.3)}] For all $\vep>0$ and $0<c<\iny$, there is a $0<b<\iny$ and $N$ -- both depending on $\vep$ and $c$ -- such that
  \benn
    \mathbb{P}\Big( \inf_{\bth\notin \cN_{b}(\bth_{0})} |\cL(\bth)|> c \Big)\geq 1-\vep, \qquad \textrm{ for all }n\geq N.
  \eenn
\end{itemize}
Note that \tbf{(u.1)} implies that $\cL$ can be approximated by some quadratic function, while the other two conditions imply that it will explode outside the neighborhood, thereby ensuring the existence of the optimal solution -- that is, the MD estimator -- inside the neighborhood. The next lemma is a reproduced version of Theorem 5.4.1 from \cite{Koul2002}, which enables this study to obtain the asymptotic normality of the MD estimator.
\begin{lem}\label{lem:asym_distr}
Suppose \textbf{(u.1)}-\textbf{(u.3)} hold. Let $\widehat{\bth}$ denote the MD estimator that solves the optimization problem in (\ref{eq:opt}). Then, there exists
$\D_{n}$ such that
\benn
\D_{n}(\widehat{\bth}-\bth_{0}) = -\left[\D_{n}\bW_{n}^{-1}(\bth_{0})\D_{n}\right]\left[\D_{n}^{-1}\bSn(\bth_{0})\right]+o_{p}(1).
\eenn
\end{lem}
In the next section, we will verify the asymptotic normality of $\bSn(\bth_{0})$, and hence, that of the MD estimator. To this end, the following assumptions are required; these assumptions have a root in \cite{Koul2002} and \cite{Kim2026}.
\be
\item[(\tbf{a.\addtocounter{Qcounter}{1}\theQcounter})] For $d_{ni}\in \mathbb{R},\,1\le i\le n$, $\sti d_{ni}^{2}=1$ and $\max_{1\le i \le n}d_{ni}^{2}=o(1)$.
\item[(\tbf{a.\addtocounter{Qcounter}{1}\theQcounter})]Let $\bar{d}_{n}:=n^{-1}\sti d_{ni}$. For any unit vector $\mbf{e}\in \mR^2$,  $\|n\bar{d}_{n}\D_{n}\mbf{e}\|=O(1)$.
\item[(\tbf{a.\addtocounter{Qcounter}{1}\theQcounter})] Let $\bg(\cdot,\bth)=\partial F(\cdot;\bth)/\partial \bth$. Then
    \benn
    \int \|\bg(y,\bth)\|^2 dH(y)=O(1),
    \eenn
    and
    \benn
    \sup \int \|\bg(y,\bth)-\bg(y,\bth_{0})\|^2 dH(y)=o(1),
    \eenn
    where the supremum is taken over $\bth\in\cN_{b}(\bth_{0})$.
\item[(\tbf{a.\addtocounter{Qcounter}{1}\theQcounter})] For $\bth\in \mathbb{R}^{2}$,
    \benn
    \int F(y;\bth)(1-F(y;\bth))dH(y)=O(1).
    \eenn
\ee
\begin{rem}
If $d_{ni}=1/\sqrt{n},\,1\le i\le n$ are used, \tbf{(a.1)} will be easily met. For the uniform $d_{ni}$, $\D_{n}=\sqrt{n}\mathbf{I}_{2\times 2}$ -- where $\mathbf{I}_{2\times 2}$ is a $2\times 2$ identity matrix -- will satisfy \tbf{(a.2)}.
\end{rem}
\begin{rem}
It is worth mentioning that the assumptions \tbf{(a.3)} and \tbf{(a.4)} will be satisfied by many location-scale family members. It turns out that the partial derivatives of $F$ in (\ref{eq:partial_derivative_F}) play an important role. Consider a normal distribution with $f$ and $F$. Let $\bg(y,\bth)=(g_{1}, g_{2})$, that is, $g_{1}(y,\bth)=-f(y;\bth)$ and $g_{2}(y,\bth)=-(y-\mu)/\si f(y;\bth)$. Consider $\bth\in \cN_{b}(\bth_{0})$. More specifically, for $\bu=(u_{1}, u_{2})\in \mR^2$, let $\D^{-1}\bu=(\bth-\bth_{0})$ where $\D_{n}=\sqrt{n}\mathbf{I}_{2\times 2}$ with $\|\bu\|\le b$. Thus, the first claim of \tbf{(a.3)} will be met if
\benn
\sup_{\|\bu\|\le b} \int |g_{i}(y,\bth_{n})-g_{i}(y,\bth_{0})|^2 dH(y)=o(1), \quad \textrm{ for }i=1,2.
\eenn
Consider $g_{2}$ and $H(y)\equiv y$; the condition will be simply met for any bounded integrating measure. Let $\mbf{q}(y,\bth)=\partial g_{2}(y,\bth)/\partial \bth$. Observe that \benn
\frac{\partial f(y;\bth)}{\partial \mu}=\frac{(y-\mu)}{\si^2}f(y;\bth),\quad
\frac{\partial f(y;\bth)}{\partial \si}=\left\{\frac{1}{\si}+\left(\frac{y-\mu}{\si}\right)^2\right\}f(y;\bth)
\eenn

Using this, another direct, albeit long, calculation shows that
\benrr
\|\mbf{q}(y,\bth)\|^2 &\le &2\left|\frac{\partial}{\partial \mu}\left(\frac{(y-\mu)}{\si}f(y;\bth)\right)\right|^2+2\left|\frac{\partial}{\partial \si}\left(\frac{(y-\mu)}{\si}f(y;\bth)\right)\right|^2,\\
&\le& \frac{4}{\si^2}f^2(y;\bth) + \frac{4}{\si^2}\left(\frac{y-\mu}{\si}\right)^4f^2(y;\bth) \\
&&\quad + \frac{8}{\si^2}\left(\frac{y-\mu}{\si}\right)^2f^2(y;\bth)+4\left(\frac{y-\mu}{\si}\right)^6f^2(y;\bth),
\eenrr
where both inequalities follow from the fact that $(a+b)^2\le 2(a^2+b^2)$ for $a,b\in \mR$.
Note that for any finite, even integer $k=2m$,
\benn
\int_{-\iny}^{\iny} \left(\frac{y-\mu}{\si}\right)^k f^2(y;\bth)dy \propto \mE(Z^{k}) =  C(2m-1)\times\cdots 3\times 1<\iny,
\eenn
where $Z$ is a standard normal random variable. In view of this, $\int \|\mbf{q}(y,\bth)\|^2\, dy<\iny$ is crystal-clear, and hence, we have
\benrr
\int |g_{2}(y,\bth)-g_{2}(y,\bth_{0})|^2\,dy &=&\int |(\bth-\bth_{0})'\mbf{q}(y,\wt{\bth})|^{2}\,dy,\\
&\le & \int \|\bth-\bth_{0}\|^2\cdot \|\mbf{q}(y,\wt{\bth})\|^2\,dy,\\
&\le &bn^{-1}\int \|\mbf{q}(y,\wt{\bth})\|^2\,dy\longrightarrow 0,
\eenrr
where the equality follows from the mean value theorem with $\wt{\bth}$ lying between $\bth_{0}$ and $\bth_{n}$ while the first inequality is immediate from the Cauchy-Schwarz inequality. The verification of the claim for $g_{1}$ is similar to but much simpler than that for $g_{2}$, and hence, we don't include here. Note that $F(1-F)$ will be symmetric around $\mu$, and hence, (\tbf{a.4}) is trivial.
\end{rem}
\begin{rem}
Consider the exponential probability distribution with a scale parameter $\si$ only. Note that $F(y;\si)=1-e^{- y/\si}$ and $g(y, \si)=-(y/\si^2)e^{- y/\si}$. Thus, we have $\int F(y;\si)(1-F(y;\si))dy=\si/2$ and $\int \{g(y,\si)\}^2 dy = 1/(4\si)$, thereby showing that the claim of \tbf{(a.4)} and the first one of \tbf{(a.3)} hold true. The first claim of \tbf{(a.3)} can be shown by using the convergence of the integrand to 0 over $\cN_{b}(\si_{0})$ and the dominated convergence theorem. If some bounded integrating measure is used for $H(y)$ -- e.g., a probability measure -- then \tbf{(a.3)} and \tbf{(a.4)} will be trivially met.
\end{rem}

\subsection{MD estimator of the parameter $\bth_{0}$}\label{Sec:MDE_beta}
\begin{figure}[h]
\centering
\begin{subfigure}{0.45\textwidth}
\includegraphics[width=\textwidth]{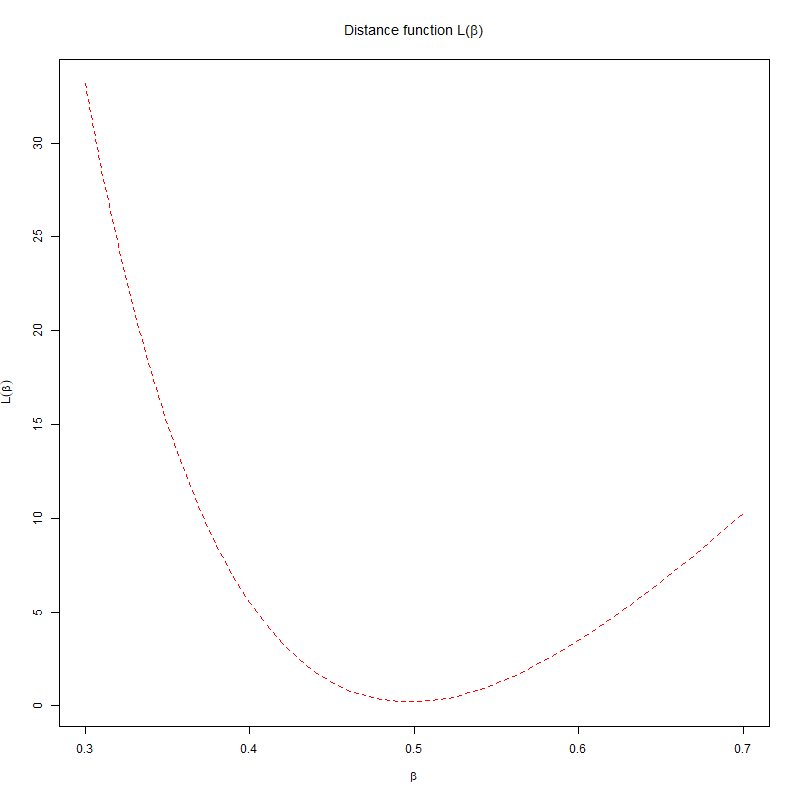}
\end{subfigure}
\begin{subfigure}{0.5\textwidth}
\includegraphics[width=\textwidth]{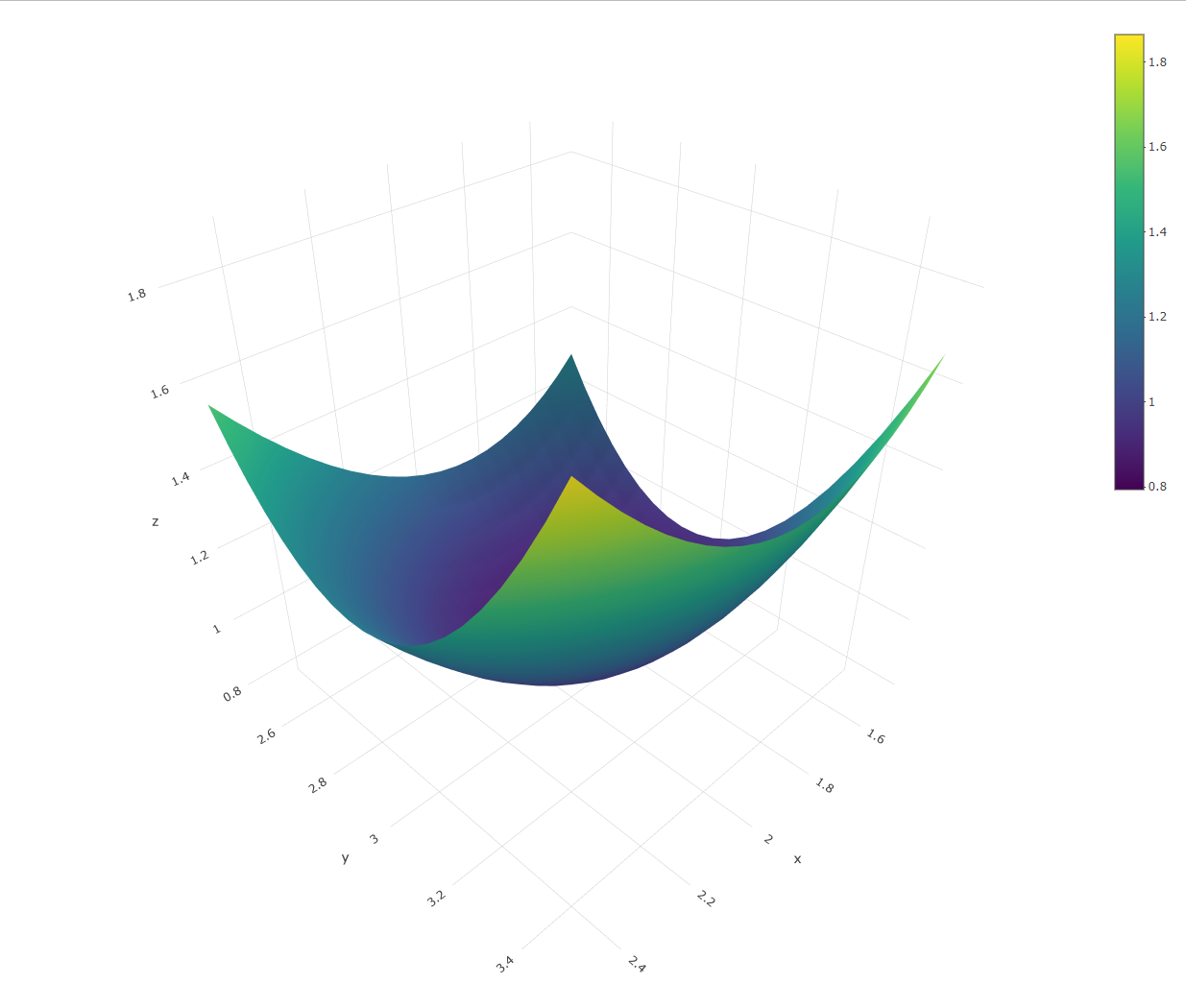}
\end{subfigure}
\caption{Graphs of distance functions from exponential (left) and normal (right) random samples}\label{fig:distant_function}
\end{figure}

Once the ULAQ conditions \textbf{(u.1)}-\textbf{(u.3)} are satisfied, Lemma \ref{lem:asym_distr} -- together with the asymptotic normality of $\D_{n}^{-1}\bSn(\bth_{0})$ and the convergence of $\D_{n}^{-1}\bW_{n}(\bth_{0})\D_{n}^{-1}$ to some quantity -- will establish the asymptotic normality of the MD estimator $\widehat{\bth}$. To begin with, define
\benn
\cW_{d}(y, \bth):= \sti d_{ni}\Big\{\textrm{I}(X_{i}\le y)-F(y;\bth) \Big\},
\eenn
and rewrite the distance function as $\cL(\bth) = \int\left[ \cW_{d}(y, \bth) \right]^{2}dH(y)$. Recall $\bg(\cdot,\bth)$ from \tbf{(a.3)}, and let
\benn
\bSn(\bth)=- n\bar{d}_{n}\int\cW_{d}(y, \bth) \bg(y,\bth)dH(y),\quad
\bW_{n}(\bth)=n^{2}\bar{d}_{n}^{2}\int \bg(y,\bth)\bg'(y,\bth)dH(y).
\eenn
Next, define
\benn
\cQ(\bth) = \cL(\bth_{0})+(\bth-\bth_{0})'\bSn(\bth_{0})+(\bth-\bth_{0})'\bW_{n}(\bth_{0})(\bth-\bth_{0}).
\eenn
Observe that $\cQ(\bth)$ can be rewritten as
$\int \left[ \cW_{d}(y, \bth_{0})-n\bar{d}_{n}(\bth-\bth_{0})'\bg(y,\bth_{0}) \right]^{2}d(y)$, which is quadratic in $(\bth-\bth_{0})$. Therefore, being able to  approximate the distance function by $\cQ(\bth)$ as $n$ increases will result in the asymptotic quadraticity of $\cL(\bth)$, which is the first ULAQ condition; its asymptotic quadraticity in $\bth$ together with smoothness plays a decisive role in reducing the computational cost of solving optimization problem in (\ref{eq:opt}). Recall $\cN_{b}(\bth_{0})=\{\bth\in\mathbb{R}^{2}:\|\D_{n}(\bth-\bth_{0})\|\le b\}$. The uniformly small difference between $\cL(\bth)$ and $\cQ(\bth)$ over $\cN_{b}(\bth_{0})$ will follow from the next theorem.
\begin{thm}\label{thm:ulaq1}
Suppose that assumptions \tbf{(a.1)}-\tbf{(a.4)} hold. Then, the distance function $\cL$ in (\ref{eq:dist_fct}) satisfies \textbf{(u.1)}, that is, for any $0<b<\iny$,
\benn
   \mathbb{E}\Big(\sup|\cL(\bth)- \cQ(\bth)|\Big)=o(1),
\eenn
where the supremum is taken over $\cN_{b}(\bth_{0})$.
\end{thm}
\begin{proof} Rewrite
$\cL(\bth) = \int \left[ \cW_{d}(y, \bth_{0}) -  n\bar{d}_{n}\{F(y;\bth)-F(y;\bth_{0}) \} \right]^{2}dH(y)$. Let $G(\bth):=F(\cdot;\bth)$ to conserve the space. Observe that
\benrr
\cL(\bth)-\cQ(\bth) &=& \int \big[ n\dn\{G(\bth)-G(\bth_{0})-(\bth-\bth_{0})'\bg(y,\bth_{0}) \} \big]^2dH(y)\\
&&-2\int \big[\cW_{d}(y, \bth_{0}) -  n\bar{d}_{n}(\bth-\bth_{0})'\bg(y,\bth_{0})\} \big] \times
\big[n\dn\{G(\bth)-G(\bth_{0})-(\bth-\bth_{0})'\bg(y;\bth_{0}) \}\big]dH(y).
\eenrr
First of all, we shall show that
\ben\label{eq:G_m_G_g}
\sup_{\bth\in \cN(\bth_{0})}\int \big[ n\dn\{G(\bth)-G(\bth_{0})-(\bth-\bth_{0})'g(y,\bth_{0}) \} \big]^2dH(y)=o(1),
\een
and
\ben\label{eq:W_m_F}
\sup_{\bth\in \cN(\bth_{0})}\int \big[\cW_{d}(y, \bth_{0}) -  n\bar{d}_{n}(\bth-\bth_{0})'\bg(y,\bth_{0})\} \big]^2dH(y)=O_{p}(1).
\een
After proving (\ref{eq:G_m_G_g}) and (\ref{eq:W_m_F}), application of the Cauchy-Schwarz inequality will complete the proof of the theorem. Recall $\bu=\D_{n}(\bth-\bth_{0})$. For $\|\bu\|\le b$, it can be rewritten as $\bu=r\mbf{e}$, where $r\le b$ and $\mbf{e}\in\mR^2$ is an unit vector. Note that
\benrr
\sup_{\|u\|\le b}\int \Big[n\bar{d}_{n}\{G(\bth)-G(\bth_{0})-(\bth-\bth_{0})'\bg(y,\bth_{0}) \}  \Big]^{2}dH(y)
&\le& b^2\|n\bar{d}_{n}\D_{n}\mbf{e}\|^2\sup_{\|u\|\le b}\int \|\bg(y,\wt{\bth})-\bg(y,\bth_{0})\|^2dH(y),\\
&&\longrightarrow 0,
\eenrr
where $\wt{\bth}\in \cN(\bth_{0})$, and the convergence to 0 follows from (\tbf{a.2}) and (\tbf{a.3}), thereby completing the proof of (\ref{eq:G_m_G_g}). To prove (\ref{eq:W_m_F}), it suffices to show that
\benn
\mathbb{E}\int \Big[ \cW_{d}(y, \bth_{0}) \Big]^2 dH(y)=O(1),\quad \mathbb{E}\sup_{\|u\|\leq b} \int \Big[ n\bar{d}_{n}(\bth-\bth_{0})'\bg(y,\bth_{0}) \Big]^2 dH(y)=O(1).
\eenn
The first equation is immediate from (\tbf{a.4}), while the second one follows from (\tbf{a.2}) and (\tbf{a.3}) after replacing $(\bth-\bth_{0})$ by $\D_{n}^{-1}\bu$.
\end{proof}
Figure \ref{fig:distant_function} shows the graphs of $\cL$ when 
a random sample of 100 observations is generated from an exponential distribution (left) with $\si=0.5$ and a normal distribution (right) with $(\mu,\si)=(2,3)$. As illustrated in the figure, $\cL$'s are quadratic functions of the parameter and attains its minimum around the true parameter, thereby empirically supporting the claim of Theorem \ref{thm:ulaq1}.

\begin{lem}\label{lem:asym_bound}
Suppose the assumptions of Theorem \ref{thm:ulaq1} hold. Then, \textbf{(u.2)} and \textbf{(u.3)} hold true.
\end{lem}
\begin{rem}
The proof of the lemma will be almost the same as -- but much simpler than -- that of Lemma 5.5.4 from \cite{Koul2002}, and hence, we do not include it here.
\end{rem}
\noi
In Lemma \ref{lem:asymp_distr_S} below, we will prove the asymptotic normality of $\bSn$ and the convergence of $\bW_{n}$. Once we prove the claims, then Lemma \ref{lem:asym_distr} will imply the asymptotic normality of the MD estimator. To this end, we need the following variable. Let $\bPsi(x,\bth):=(\psi_{1}(x,\bth),\psi_{2}(x,\bth))'$ where $\psi_{i}(x,\bth):=\int_{x}^{\iny}g_{i}(y,\bth)dH(y),\,i=1,2$. The application of Fubini's theorem yields
\benn
\mathbb{E}\bPsi(X,\bth_{0})= \int f(x)\int_{x}^{\iny}\bg(y,\bth_{0})dH(y)\, dx
= \int F(y)\bg(y,\bth_{0})dH(y).
\eenn
Recall $\bg(y,\bth)=(-f(y;\bth),-(y-\mu)/\si f(y;\bth))'$ and consider $H(y)\equiv y$. A direct calculation shows that $\psi_{1}(X_{i},\bth)=-1+F(X_{i};\bth)$, and hence, $\mE[\psi_{1}(X_{i},\bth)]=-1/2$, regardless of probability distributions. In addition, $\mE[F(X_{i};\bth)]^2=1/3$, and hence, $Var(\psi_{1}(X_{i},\bth))=1/12$. If a different $H$ is used, $\mE\psi_{1}$, and hence, $Var(\psi_{1})$, will take different values. In the literature on MD estimation theories, other popular integrating measures for $H$ include, a probability measure, Dirac delta measure, etc.

Unlike $\psi_{1}$ and its expectation, $\psi_{2}$ and $\mE\psi_{2}$ take a bit more complicated forms, depending on both $H(y)$ and $F$. Consider $H(y)\equiv y$ and a normal $F$. Then, we have
\benn
\psi_{2}(X_{i},\bth)=-\int_{(X_{i}-\mu)/\si}^{\iny}\frac{1}{\sqrt{2\pi}}xe^{-x^2/2}dx=-\frac{1}{\sqrt{2\pi}}e^{-(X_{i}-\mu)^2/2\si^2}.
\eenn
Consequently, it can be shown that $\mE [\psi_{2}(X_{i},\bth)]=-1/(2\sqrt{\pi})$ and $Var[\psi_{2}(X_{i},\bth)]=(12^{-1/2}-0.25)/\pi\approx 0.012$. Finally, rewrite $\bPsi$ and $\mE\bPsi$ as
\benn
\bPsi(X_{i},\bth)=\left(\begin{array}{c}
                -1+F(X_{i};\bth) \\
                -\si f(X_{i};\bth)
              \end{array}\right),\quad
\mE[\bPsi(X_{i},\bth)]=\left(\begin{array}{c}
                -1/2 \\
                -1/(2\sqrt{\pi})
              \end{array}\right).
\eenn
The fact that both $\bPsi$ and $\mE\bPsi$ have analytic expressions as in the above equations facilitates the fast computation of the MD estimator for a normal case since $\bSn$ -- which plays the most important role when we seek the estimator -- is written as those two variables. Similar fact holds for many other probability distributions of a location-scale family, including logistic, Cauchy, exponential, and Weibull distributions.

Define a covariance matrix $\bO(\bth_{0}):=Var(\bPsi(X,\bth_{0}))$, that is,
\benn
\bO(\bth_{0})= \left(
        \begin{array}{cc}
          Var(\psi_{1}(X,\bth_{0})) & Cov(\psi_{1}(X,\bth_{0}), \psi_{2}(X,\bth_{0})) \\
          Cov(\psi_{1}(X,\bth_{0}), \psi_{2}(X,\bth_{0})) & Var(\psi_{2}(X,\bth_{0})) \\
        \end{array}
      \right).
\eenn

\begin{lem}\label{lem:asymp_distr_S}
Suppose the assumptions of Theorem \ref{thm:ulaq1} continue to hold. Let  $\bW(\bth):=\int \bg(y,\bth)\bg'(y,\bth)dH(y)$. Then the following hold true:
\benn
\lim_{n\rightarrow \iny}\D_{n}^{-1}\bW_{n}(\bth_{0})\D_{n}^{-1}=\bW(\bth_{0}).
\eenn
and
\benn
\D_{n}^{-1}\bSn(\bth_{0})\Rightarrow_{\cD}N(\mbf{0}, \bO(\bth_{0})),
\eenn
where $\mbf{0}$ is a $2\times 1$ vector of zeros.

\end{lem}
\noi
\begin{proof}
Convergence of $\D_{n}^{-1}\bW_{n}(\b_{0})\D_{n}^{-1}$ is straightforward from (\tbf{a.2}). Note that $\D_{n}^{-1}\bSn(\bth_{0})=-(n\bar{d}_{n}\D_{n}^{-1})\mbf{T}_{n}$
where
\benn
\mbf{T}_{n} = \int \cW_{d}(y, \bth_{0})\bg(y,\bth_{0})dH(y).
\eenn
Next, define $\bE_{k}(\bth_{0}):=(\eta_{k1}(\bth_{0}),\eta_{k2}(\bth_{0}))'\in \mR^2,\,1\le k\le n$, where
\benn
\eta_{ki}(\bth_{0}):=\int g_{i}(y,\bth_{0})\Big\{\textrm{I}(X_{k}\le y)-F(y;\bth_{0})\Big\}dH(y),\quad 1\le i\le 2.
\eenn
Subsequently, rewrite $\mbf{T}_{n}=\stk d_{nk}\bE_{k}$.
and hence, $\bE_{k}$ can be expressed as $\bE_{k} = \bPsi(X_{k},\bth_{0})-\mE\bPsi(X_{k},\bth_{0})$, implying $\mE(\bE_{k})=0,\,1\le k\le n$.

Consider $\mbf{b}\in \mR^{2}$. Let $\tau_{n}^2:= \stk d_{nk}^{2}Var(\mbf{b}'\bE_{k})$. Since $\bO(\bth_{0})=\mE[\bE_{1}(\bth_{0})\bE_{1}'(\bth_{0})]$, the iid assumption and \tbf{(a.1)} imply $\tau_{n}^2=\mbf{b}'\bO \mbf{b}$, and the first equation of \tbf{(a.3)} implies $|\mbf{b}'\bE_{k}|<\iny$. Then, for any $\vep>0$
\benrr
\tau_{n}^{-2}\sti d_{ni}^{2} \mathbb{E}\left[|\mbf{b}'\bE_{k}|^{2}: d_{ni}|\mbf{b}'\bE_{k}|\geq \vep \tau_{n}\right]&\leq & C\tau_{n}^{-2} \max_{1\le i\le n}d_{ni}^2\sti \mathbb{P}(d_{ni}|\mbf{b}'\bE_{k}|\geq \vep \tau_{n}),\\
&\leq & C \tau_{n}^{-2}\vep^{-2} \max_{1\le i\le n}d_{ni}^2\longrightarrow 0,
\eenrr
where the second inequality follows from Chevyshev's inequality while \tbf{(a.1)} implies the convergence to 0. Thus, Lindeberg's condition is met, and the Cramer-Wold device together with (\tbf{a.1}) yields
\benn
\mbf{T}_{n}\Rightarrow_{\cD}N(0, \bO(\bth_{0})).
\eenn
Consequently, the asymptotic convergence of $\D_{n}^{-1}\bSn(\bth_{0})$ to $N(0, \bO(\bth_{0}))$ in distribution directly follows from (\tbf{a.2}).
\end{proof}
\noi
Finally, the next theorem demonstrates the asymptotic normality of the MD estimator.
\begin{thm}\lel{thm:asymp}
Suppose the assumptions of Theorem \ref{thm:ulaq1} continue to hold, and  let $\bO(\bth)$ and $\bW(\bth)$ be as in Lemma \ref{lem:asymp_distr_S}. Define $\S(\bth):=\bW^{-1}\bO\bW^{-1}$. Then the MD estimator $\widehat{\b}$ in (\ref{eq:opt}) will be asymptotically normally distributed, that is,
\benn
\D_{n}(\wh{\bth}-\bth_{0})\Rightarrow_{\cD}N(\mbf{0},\S(\bth_{0})).
\eenn
\end{thm}
\begin{proof} Note that the ULAQ conditions are met by
Theorem \ref{thm:ulaq1} and Lemma \r{lem:asym_bound}, and hence, Lemma \ref{lem:asym_distr} accompanied by Lemma \ref{lem:asymp_distr_S} will immediately imply the claim, thereby completing the proof of the theorem.
\end{proof}
\noi
\section{Some desriable properties}\label{sec:desirable_propoerties}
We will discuss desirable properties of the MD estimator, including the efficiency and robustness.
\subsection{Efficiency of the MD estimator}
We further examine the asymptotic variance of the MD estimator for  probability distributions of a location-scale family. Let $\wh{\bth}:=(\wh{\mu},\wh{\si})'$ denote the MD estimators of the location and scale parameters, respectively. To begin with, we investigate a normal $F$. Consider $H(y)\equiv y$ again. Recall $\S(\bth)=\bW^{-1}\bO\bW^{-1}$. It is obvious that the asymptotic variance of both location and scale parameters will be determined by $Var(\bPsi)$ and $\bW$. A direct calculation shows that
\benn
\bW(\bth_{0})=\left(
                     \begin{array}{cc}
                       \int f^2(y;\bth_{0})dy & \int \frac{(y-\mu)}{\si}f^2(y;\bth_{0}) dy\\
                       \int \frac{(y-\mu)}{\si}f^2(y;\bth_{0})dy &
                       \int \frac{(y-\mu)^2}{\si^2}f^2(y;\bth_{0})dy \\
                     \end{array}
                   \right)=\left(
                     \begin{array}{cc}
                        1/(2\si\sqrt{\pi})& 0\\
                       0 & 1/(4\si\sqrt{\pi}) \\
                     \end{array}
                   \right),
\eenn
and hence, the asymptotic covariance matrix of the MD estimator will be
\benn
\S(\bth_{0})=4\pi \si_{0}^{2}\left(
        \begin{array}{cc}
          Var(\psi_{1}(X,\bth_{0})) & 2Cov(\psi_{1}(X,\bth_{0}), \psi_{2}(X,\bth_{0})) \\
          2Cov(\psi_{1}(X,\bth_{0}), \psi_{2}(X,\bth_{0})) & 4Var(\psi_{2}(X,\bth_{0})) \\
        \end{array}
      \right).
\eenn
Recall $Var(\psi_{1}(X,\bth_{0}))=1/12$ and $Var(\psi_{2}(X,\bth_{0}))\approx0.0123$. In view of this, the asymptotic variance of the MD estimator $\wh{\mu}$ is $4\pi \si^{2}/12\approx1.0472\si_{0}^2$ while that of $\wh{\si}$ will be approximately $0.618\si_{0}^2$. Given that the asymptotic variance of the ML estimator for $\mu_{0}$, the MD estimator of the location parameter is slightly less efficient than the ML estimator.

Next, we proceed to investigate the logistic distribution whose $f(\cdot;\bth)$ and $F(\cdot;\bth)$ are
\benn
f(y;\bth) = \frac{e^{-(y-\mu)/\si}}{\si(1+e^{-(y-\mu)/\si})^2},\quad F(y;\bth) = \frac{1}{1+e^{-(y-\mu)/\si}}.
\eenn
Recall that variables related with $\psi_{1}$ have the common expression, regardless of probability distributions, that is,
$\psi_{1}(X,\bth)=-1+F(X;\bth)$, $E[\psi_{1}(X,\bth)]=1/2$, $Var[\psi_{1}(X,\bth)]=1/12$, etc. Rewrite $\bW(\bth)=[[w_{ij}(\bth)]],\,1\le i,j\le 2$. Then the diagonal elements of $\bW$ -- which are required to obtain the asymptotic variance of the MD estimator -- will be $w_{11}(\bth)=1/(6\si)$ and $w_{22}(\bth)\approx0.215/\si$. Consequently , the asymptotic variance of $\wh{\mu}$, which is the MD estimator of the location parameter, will be $Var[\psi_{1}(X,\bth_{0})]/w_{11}^{2}=3\si_{0}^2$. Like the normal case, $\psi_{2}$ and its moments are much harder and longer to compute than their $\psi_{1}$'s counterparts. For example, $\psi_{2}(X;\bth)=\kappa((X-\mu)/\si)$ where $\kappa:\mR\ra\mR$ and
\benn
\kappa(y) = \frac{-ye^{-y}-(1+e^{-y})\ln(1+e^{-y}) }{1+e^{-y}}.
\eenn
Using numerical approximation, we have $\mE[\psi_{2}(X;\bth)]\approx-0.5$ and $Var[\psi_{2}(X;\bth)]\approx0.035$. Using this result, the asymptotic variance of $\wh{\si}$ will be $Var[\psi_{2}(X;\bth)]/w_{22}^{2}\approx 0.757\si^2$.

Next, consider a probability distribution with a single scale parameter, an exponential distribution whose $f(\cdot;\si)$ and $F(\cdot;\si)$ are
\benn
f(y;\si)=\frac{1}{\si}e^{-y/\si},\quad F(y;\si)=1-e^{-y/\si},
\eenn
respectively. In the exponential case, finding the asymptotic variance is much simpler, and the asymptotic variance of the MD estimator turns out to be $16\si^2(17/27-9/16)\approx1.074\si^2$. For choosing a probability measure, e.g., $H(y)=1-e^{-y}$, the MD estimator will have a larger asymptotic variance ($\approx1.314\si^2$).

\subsection{Robustness of the MD estimator}\label{Sec:IF}
This section will discuss the robustness of the MD and other estimators. For the shake of simplicity, we will consider an exponential probability distribution with a scale parameter $\si$ as in the previous section. 

\begin{figure}[h]
\centering
\begin{subfigure}{0.45\textwidth}
\includegraphics[width=\textwidth]{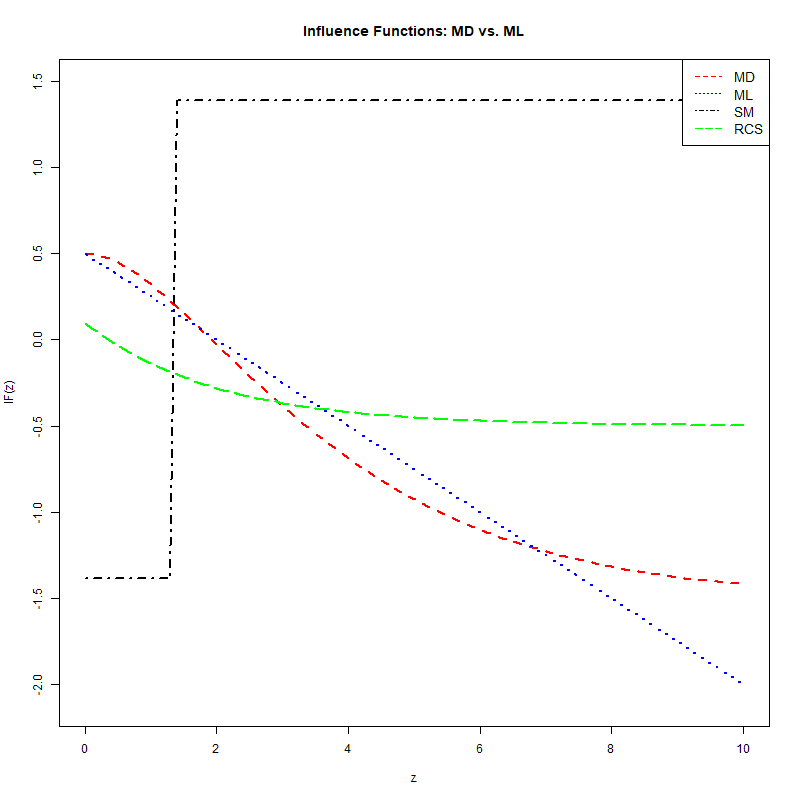}
\end{subfigure}
\begin{subfigure}{0.45\textwidth}
\includegraphics[width=\textwidth]{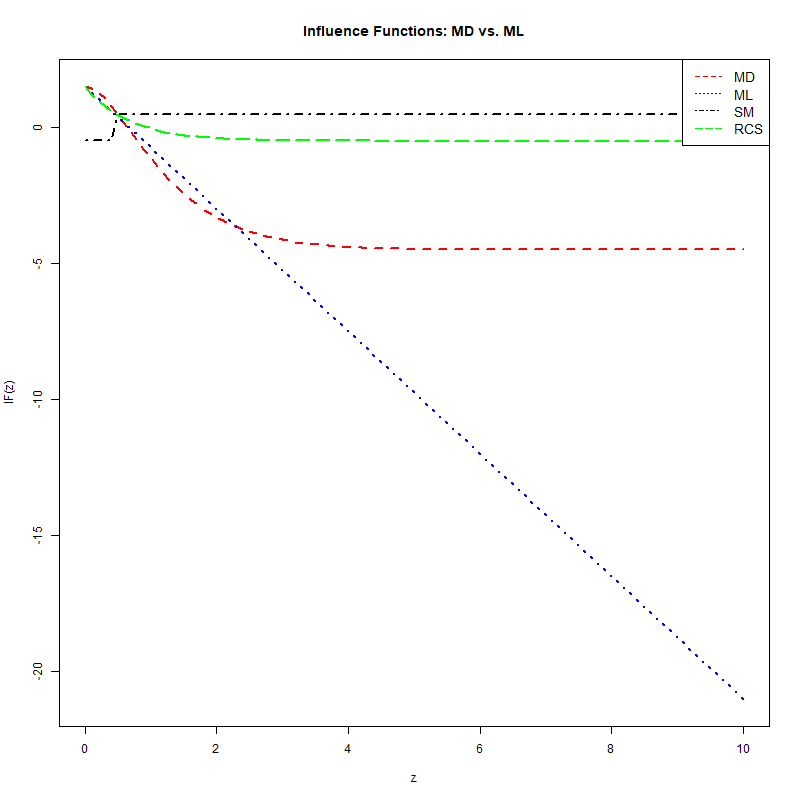}
\end{subfigure}
\caption{Influence function.}\label{fig:IF_All}
\end{figure}

When assessing the robustness of the estimation method, researchers use many measures; among them, the most popular one is the influence function proposed by \cite{Hampel1968} and \cite{Hampel1986}. When computing influence functions of the chosen estimators, we shall use the direct formula (2.3.5) in \cite{Hampel1986}. Now we are ready to find the influence function of the MD estimator. Consider the contaminated exponential probability distribution
\ben\label{eq:contaminated}
F_{\vep}(x;\si) = (1-\vep)F(x;\si) + \vep \delta_{z},
\een
where $z>0$ is an extreme value. Let $\textrm{IF}_{D}(z;\si)$ denote the influence function of the MD estimator. Note that finding the MD estimator is equivalent to solving
\benn
\sti \phi(X_{i},\si) = 0,
\eenn
where $\phi(X_{i},\si):=d_{ni} \int g(y,\si)\{I(X_{i}\le y)-F(y;\si)\}dH(y)$. Using $d_{ni}=1/\sqrt{n}$ will yield
\benn
\phi(z,\si)=  \frac{1}{\sqrt{n}}\left( \si z e^{- z/\si}+\si^2e^{- z/\si}-\frac{3\si^2}{4} \right), \qquad \mathbb{E}\left[\frac{\partial \phi(X_{i},\si)}{\partial \si}\right]=-\frac{\si^3}{4\sqrt{n}}.
\eenn
Plugging the above equations to the formula (2.3.5) in \cite{Hampel1986} will yield the influence function of the MD estimator $\textrm{IF}_{D}(z;\si) = 4\si^{-2} z e^{- z/\si}+4\si^{-1} e^{- z/\si}-3\si^{-1}$. Note that $\textrm{IF}_{D}(z;\si)$ converges to $-3/\si$ as $z$ approaches $\iny$, which implies the impact of an outlier will be limited: see also Figure \ref{fig:IF_All}. Note that the influence function of the ML estimator can be obtained much more simply: $\textrm{IF}_{L}(z;\si) = \si^{-1}-\si^{-2} z$; the impact of the extreme value $z$ on the ML estimator is unlimited, while that on the MD estimator is bounded and diminishes as $z$ approaches $\iny$. Therefore, the efficiency of the ML estimator for some probability distribution (e.g., a normal distribution) will be compromised by the presence of outliers, as their impact cannot be limited: see also Figure \ref{fig:IF_All}.

To address the presence of an outlier, the sample mean in the ML estimator is replaced by the sample median, yielding the standard median (SM) estimator, $\wh{\si}_{S}:=\textrm{Med}_{1\le i\le n}\{X_{i}\}/\log 2$, where $\log 2$ is for the Fisher-consistency. By using the general formula from \cite{Staudte}, the asymptotic variance and the influence function of the SM estimator will be $2.0814\si^2$ and
\benn
\textrm{IF}_{S}(z;\si)=\left\{
        \begin{array}{ll}
          -\frac{\si}{\log 2}, & \hbox{if $z<\si/\log 2$;} \\
          0, & \hbox{if $z=\si/\log 2$;} \\
          \frac{\si}{\log 2}, & \hbox{if $z>\si/\log 2$,}
        \end{array}
      \right.
\eenn
respectively. Hence, the relative efficiency of the SM estimator is less than 0.5, being much worse than that of the MD estimator. The influence function will be bounded above and below by $\pm \si/\log 2$. Thus, for any $\si<1.442$ -- which solves for $3/\si<\si/\log 2$ -- the impact of any outlier on the MD estimator will be smaller than that on the SM estimator, while the opposite will be true for $\si>1.442$.

Another estimator of $\si$ is the S-estimator proposed by \cite{Croux1993}. \cite{Gather1999} refer to it as the \textit{RCS} estimator, emphasizing that it differs from the general class of the S-estimator. In this study, we also use the same name as in \cite{Gather1999}. The RCS estimator, denoted by $\si_{R}$, is defined as
\benn
\wh{\si}_{R} = c\,\textrm{Med}_{1\le i\le n}\{Y_{i}\},
\eenn
where $Y_{i}:=\textrm{Med}_{1\le j\le n}\{|X_{i}-X_{j}|\}$. To ensure Fisher-consistency, $c$ takes different values depending on the probability distribution. \cite{Croux1993} discussed the RCS estimator regarding various non-Gaussian distributions, including the exponential distribution. For the exponential distribution, they demonstrated that $c$ is approximately $1.6982$, and its asymptotic variance is $1.8217\si^{-2}$; hence, the relative efficiency of the RCS estimator is 1/1.8217 ($\approx 0.547$), which is slightly better than that of the SM estimator but still much worse than that of the MD estimator. Thus, in terms of efficiency,  the MD estimator outperforms the SM and RCS estimators, except for the ML estimator.

Using Theorem 3 from \cite{Croux1993}, the influence function of the RCS estimator, $\textrm{IF}_{R}(z;\si)$, can be directly computed. However, it is too long, and hence, is not included here. Figure \ref{fig:IF_All} plots the influence functions of four estimators: 0.5 and 1.5 for $\si$ are used for the left and right figures, respectively.
As shown in the figure, regardless of $\si$, the RCS estimator performs best, while the ML estimator is the least robust. Following the RCS estimator, the MD and SM estimators vie for superiority. As mentioned before, the superiority of the MD estimator to the SM estimator will depend on $\si$: for smaller $\si$, the MD estimator outperforms the SM estimator, while the opposite holds for larger $\si$.

\section{Simulation studies}\label{Sec:simulation_studies}
We generate $n$ random observations from an exponential distribution with the true rate parameter of $0.8$. Using the generated sample, we compare the performance of ML, SM, RCS, and MD estimators of the previous section. For comparison purposes, we will use bias, standard error (SE), and root mean square error (RMSE) as evaluation criteria.
\begin{table}[h!]
\centering
\caption{Bias (left), SE (middle), and RMSE (right) of estimators with no outlier.}\label{tbl:RMSE}
\begin{tabular}{|c|cccc |cccc |cccc|}
\cline{1-13}
 & \multicolumn{4}{c}{Bias} & \multicolumn{4}{|c}{SE} & \multicolumn{4}{|c|}{RMSE}\\
\cline{1-13}
  $n$ & ML &  SM & RCS & MD & ML &  SM & RCS & MD & ML &  SM & RCS & MD \\
\cline{1-13}
20   &0.042 &0.052 &0.137 &0.038 & 0.194 &0.285 &0.315 &0.187 & 0.199 &0.29  &0.344 &0.191\\
\cline{1-13}
40   &0.019 &0.025 &0.064 &0.018 & 0.131 &0.19  &0.196 &0.133 & 0.132 &0.192 &0.206 &0.134\\
\cline{1-13}
60   &0.014 &0.019 &0.044 &0.013 & 0.106 &0.157 &0.154 &0.11 & 0.107 &0.158 &0.16  &0.111\\
\cline{1-13}
80   &0.01  &0.012 &0.03  &0.009 & 0.092 &0.132 &0.129 &0.095 & 0.092 &0.132 &0.132 &0.096\\
\cline{1-13}
100  &0.009 &0.012 &0.027 &0.007 & 0.082 &0.118 &0.116 &0.086 & 0.082 &0.119 &0.119 &0.087\\
\cline{1-13}
\end{tabular}
\end{table}
Table \ref{tbl:RMSE} reports bias, SE, and RMSE of four estimators. First of all, all estimators show consistency; as $n$ increases, their biases decrease. The MD estimator yields the smallest bias, followed by the ML estimator, which shows slightly larger bias, and the SM estimator. Note that the RCS estimator performs worst; regardless of $n$, it always exhibits much larger bias relative to others. Conversely, the ML estimator shows slightly better SE than the MD estimator; these two estimators again outperform the SM and RCS estimators, as their SEs are much smaller than those of their competitors, which closely accords with the findings in Section \ref{Sec:IF}.

In conclusion, the ML and MD estimators exhibit nearly identical performance, outperforming others by maintaining the lowest bias or SE across all $n$'s. Among all, the RCS estimator reports the worst overall performance. After exhibiting the largest bias (0.137), SE (0.315), and RMSE (0.344) at $n=20$, it remains worse than the ML and MD estimators for both bias and SE until $n$ reaches 100. The SM estimator remains middle-ranked between the top-tier group (ML and MD) and the RCS estimator across all $n's$. As $n$ increases, the SM estimator exhibits the diminishing superiority to RCS in RMSE (both RMSEs converging to 0.119), while showing similar SE's to and smaller biases than RCS.

For the next experiment, we generate random observations from the contaminated exponential distribution in (\ref{eq:contaminated}) where $\eps=0.01$ and $z=10$. Table \ref{tbl:RMSE_z=10} reports the results in the presence of outliers.
\begin{table}[h]
\centering
\caption{Bias (left), SE (middle), and RMSE (right) of estimators with an outlier $z=10$.}\label{tbl:RMSE_z=10}
\begin{tabular}{|c|cccc |cccc |cccc| }
\cline{1-13}
 & \multicolumn{4}{c}{Bias} & \multicolumn{4}{|c}{SE} & \multicolumn{4}{|c|}{RMSE}\\
\cline{1-13}
  $n$ & ML &  SM & RCS & MD & ML &  SM & RCS & MD & ML &  SM & RCS & MD \\
\cline{1-13}
20   &-0.002 &0.04  &0.12  &0.016 &  0.204 &0.283 &0.311 &0.191 & 0.204 &0.286 &0.333 &0.191\\
\cline{1-13}
40   &-0.027 &0.018 &0.052 &-0.003 & 0.139 &0.195 &0.197 &0.137 & 0.142 &0.196 &0.204 &0.137\\
\cline{1-13}
60   &-0.036 &0.008 &0.029 &-0.011 & 0.113 &0.153 &0.154 &0.111  & 0.119 &0.153 &0.157 &0.112\\
\cline{1-13}
80   &-0.04  &0.003 &0.018 &-0.015 & 0.098 &0.134 &0.133 &0.098 & 0.106 &0.134 &0.134 &0.099\\
\cline{1-13}
100  &-0.042 &0.002 &0.013 &-0.016 & 0.087 &0.118 &0.116 &0.089 & 0.096 &0.118 &0.116 &0.09\\
\cline{1-13}
\end{tabular}
\end{table}
As in the previous experiment, decreases in SE lead to decreases in RMSE across all estimation methods, as the magnitudes of their biases are relatively small and cannot have a significant impact on corresponding RMSEs. Also, the ML and MD methods again outperform the other two methods in terms of RMSE; the MD estimator takes the lead as the best one, uniformly exhibiting the smallest RMSE, while the RCS estimator still reports the worst RMSE for all $n$ except for 100. It is worth noting that only the SM and RCS estimators exhibit asymptotic consistency regarding bias; the biases of both methods decrease as $n$ increases. As $n$ increases, the magnitude of bias of the ML estimator monotonically increases, whereas that of the MD estimator fluctuates and is stabilized at 0.016.

From the results reported in Tables \ref{tbl:RMSE} and \ref{tbl:RMSE_z=10}, the ML and MD estimators outperform the other two estimators, regardless of the presence of an outlier. While the MD estimator exhibits a slightly smaller bias than the ML estimator, it has a larger SE, leading to a larger RMSE than its competitor. However, the presence of an outlier tilts the balance in favor of the MD estimator. For most $n$'s, the MD estimator exhibits smaller bias and SE, resulting in a smaller RMSE than the ML estimator, since the ML estimator can not be free from the impact of the outlier. Thus, it is not rash to conclude that in the presence of an outlier, the MD estimator is strongly recommended among all estimators.

\section{Conclusion}\label{Sec:conclusion}
This study applied the MD estimation with the CvM-type distance function, along with a different approach proposed by \cite{Kim2026}, to estimate parameters of various probability distributions belonging to a location and scale family and demonstrated that the MD estimator still retains desirable properties, such as asymptotic normality, efficiency, and robustness, and outperforms other estimators in the presence of an outlier. Further extensions of the current study to more advanced statistical models such as survival analysis and time series analysis will form the basis for future research.


\bibliographystyle{plain}
\bibliography{MDE_ref}

\edt

outlandish, outrageous, paltry

%% file: MyMacro.tex
\usepackage{amsfonts}
\usepackage{float}
\usepackage{amsmath}
\usepackage{amssymb}
\usepackage{amsthm}
\usepackage{graphicx}
\usepackage{multirow}
\usepackage{color}
\usepackage{ulem}
\usepackage{setspace}
\usepackage{caption}
\usepackage{subcaption}
\usepackage{rotating}
\usepackage{threeparttable}
\usepackage{url}
\usepackage{framed}

\newcommand{\mbf}[1]{\mbox{\boldmath $#1$}}

\def\bu{\mbf{u}}

\def\bW{\mathbf{W}}

{\catcode `\@=11 \global\let\AddToReset=\@addtoreset}
\AddToReset{equation}{section}

\AddToReset{Theorem}{section}

\newtheorem{lem}{Lemma}[section]
\newtheorem{thm}{Theorem}[section]

\theoremstyle{remark}
\newtheorem{rem}{Remark}[section]

\newcommand{\cD}{{\cal D}}

\newcommand{\cL}{{\cal L}}

\newcommand{\cN}{{\cal N}}

\newcommand{\cQ}{{\cal Q}}

\newcommand{\cW}{{\cal W}}

\def\bc{\begin{center}}
\def\bd{\begin{description}}
\def\be{\begin{enumerate}}
\def\ec{\end{center}}
\def\ed{\end{description}}

\def\ee{\end{enumerate}}
\def\ben{\begin{equation}}
\def\benn{\begin{equation*}}
\def\een{\end{equation}}
\def\eenn{\end{equation*}}
\def\benr{\begin{eqnarray}}
\def\eenr{\end{eqnarray}}
\def\benrr{\begin{eqnarray*}}
\def\eenrr{\end{eqnarray*}}

\def\b{\beta}

\def\D{\Delta}

\def\edt{\end{document}}

\def\iny{\infty}

\def\lel{\label}

\def\noi{\noindent}

\def\r{\ref}

\def\ra{\rightarrow}

\def\si{\sigma}

\def\sti{\sum_{i=1}^n}

\def\stk{\sum_{k=0}^\iny}

\def\vep{\varepsilon}
\def\eps{\epsilon}

\def\wh{\widehat}

\def\wt{\widetilde}

\def\dn{\bar{d}_{n}}

\def\bSn{\mbf{S}_{n}}

\def\bu{\mbf{u}}

\def\bg{\mbf{g}}

\def\bth{\mbf{\theta}}

\def\D{\boldsymbol{\delta}}

\def\S{\boldsymbol{\Sigma}}

\def\bO{\boldsymbol{\Omega}}

\def\bE{\boldsymbol{\eta}}
\def\bPsi{\boldsymbol{\psi}}

\def\mR{\mathbb{R}}

\def\mE{\mathbb{E}}

\def\vX{X_{1},...,X_{n}}

\def\tbf{\textbf}

\newcounter{Qcounter}
\newcounter{show}